\documentclass[10pt]{article}
\usepackage{graphicx}
\usepackage{lineno}

\def\Title#1{\begin{center} {\Large #1 } \end{center}}
\def\Author#1{\begin{center}{ \sc #1} \end{center}}
\def\Address#1{\begin{center}{ \it #1} \end{center}}

\newcommand\pubblock{\rightline{\begin{tabular}{l} Proceedings of the Fifth Annual LHCP\\ \pubnumber\\
         \pubdate  \end{tabular}}}

\newenvironment{Abstract}{\begin{quotation} \begin{center} 
             \large ABSTRACT \end{center}\bigskip 
      \begin{center}\begin{large}}{\end{large}\end{center} \end{quotation}}

\newenvironment{Presented}{\begin{quotation} \begin{center} 
             PRESENTED AT\end{center}\bigskip 
      \begin{center}\begin{large}}{\end{large}\end{center} \end{quotation}}

\def\beq{\begin{equation}}
\def\eeq#1{\label{#1}\end{equation}}
\def\eeqn{\end{equation}}

\def\beqa{\begin{eqnarray}}
\def\eeqa#1{\label{#1}\end{eqnarray}}
\def\eeqan{\end{eqnarray}}

\let\bar=\overbar

\def\Dslash{\not{\hbox{\kern-4pt $D$}}}
\def\dslash{\not{\hbox{\kern-2pt $\del$}}}

\def\msb{{\bar{\ssstyle M \kern -1pt S}}}

\usepackage{color}

\newcommand\pubnumber{ }

\newcommand\pubdate{\today}

\def\affiliation{
On behalf of the ALICE, ATLAS, CMS, LHCb and TOTEM Collaborations, \\
 LAPP, CNRS/IN2P3 and Universit{\'e} Savoie Mont Blanc, Annecy-le-Vieux, France }

\begin{document}

\large
\begin{titlepage}
\pubblock

\vfill
\Title{  Soft QCD  }
\vfill

\Author{ \v{S}\'{a}rka Todorova-Nov\'{a}  }
\Address{\affiliation}
\vfill
\begin{Abstract}
    The talk contains an overview of recent experimental results obtained by the LHC
    experiments. The measured inclusive event properties, as well as
    various correlation phenomena, are compared with predictions of phenomenological models.  

\end{Abstract}
\vfill

\begin{Presented}
The Fifth Annual Conference\\
 on Large Hadron Collider Physics \\
Shanghai Jiao Tong University, Shanghai, China\\ 
May 15-20, 2017
\end{Presented}
\vfill
\end{titlepage}
\def\thefootnote{\fnsymbol{footnote}}
\setcounter{footnote}{0}
%

\normalsize 


\section{Introduction}

      The understanding of the bulk of the particle production at the
      LHC has a major impact on the search of new physics
      phenomena. The measurements of the inclusive cross-section, of
      the inclusive single particle spectra, and the measurements of various
      correlation phenomena serve both as a test of the predictions
      of the phenomenological models and as an input for their improvement.   
       
\section{Inclusive cross-section measurements }
\begin{figure}[htb]
\centering
\includegraphics[height=2in]{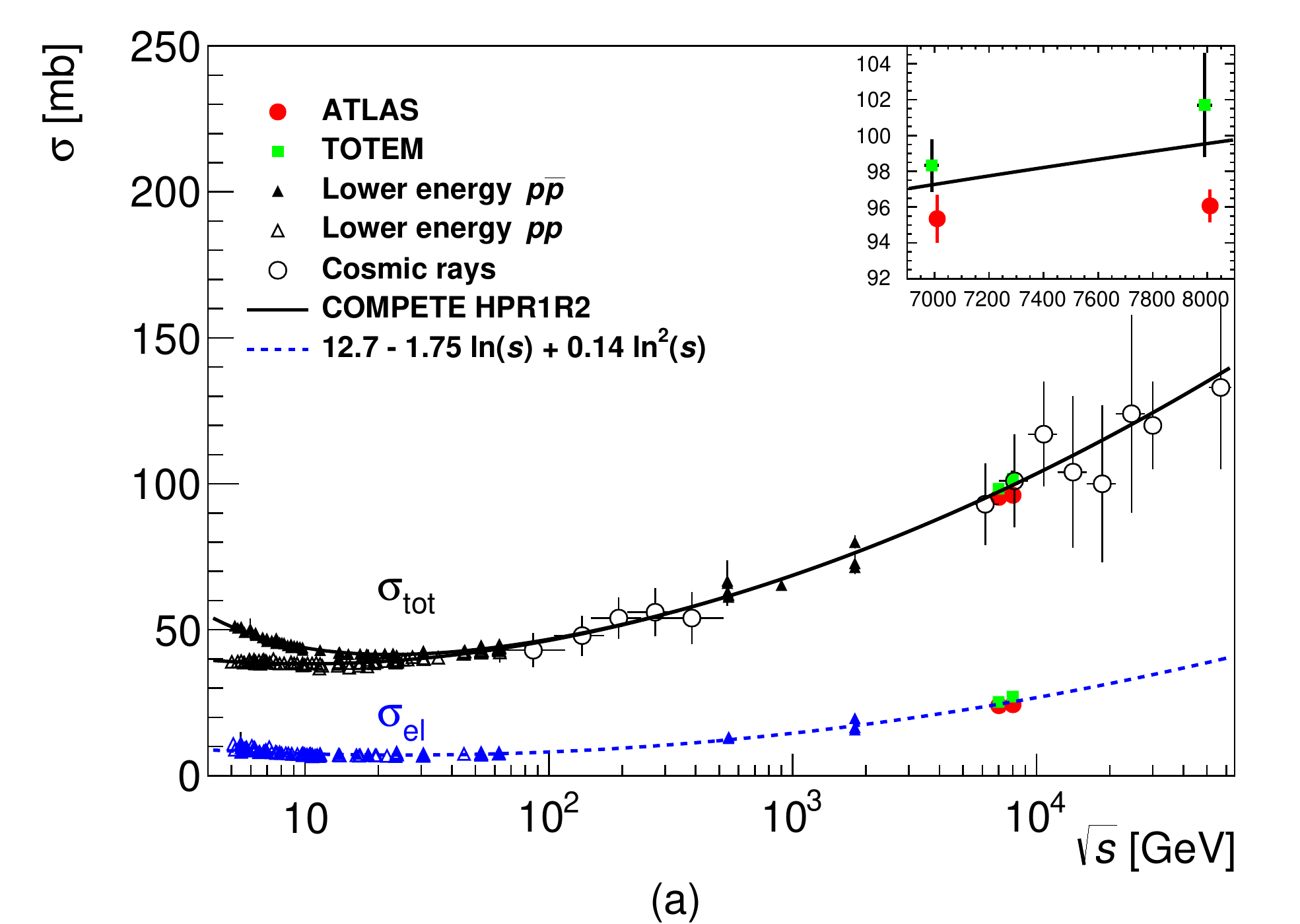}
\includegraphics[height=2in]{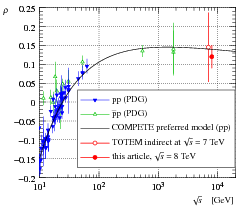}
\caption{ Left: The elastic and the total proton-proton cross sections versus
  $\sqrt{s}$~\cite{ATLAS-el}. Right: The $\rho$-parameter versus $\sqrt{s}$~\cite{TOTEM-CNI} .}
\label{fig:el}
\end{figure}
    Figure~\ref{fig:el}(left) shows the compilation of the measurements of
    the elastic and the total proton-proton cross
    section as a function of the center-of-mass energy, including the
    ATLAS and TOTEM measurements at $\sqrt{s}$=7TeV and 8TeV, compared to
    the global fit (plot taken from ~\cite{ATLAS-el}). The total
    cross section is obtained from the measured elastic scattering via
    the optical theorem $\sigma^2_{tot}=16\pi(\hbar c)^2/(1+\rho^2)d\sigma_{el}/dt|_{t->0}$. The TOTEM measurement at $\sqrt{s}$=8TeV
    uses a luminosity-independent method ~\cite{TOTEM-el}. Figure~\ref{fig:el}(right) shows the
    energy dependence of the $\rho$ parameter including the TOTEM
    measurement in the region of very low momentum transfer $t$ sensitive
    to the  Coulomb-nuclear interference~\cite{TOTEM-CNI}.
\begin{figure}[htb]
\centering
\includegraphics[height=2in]{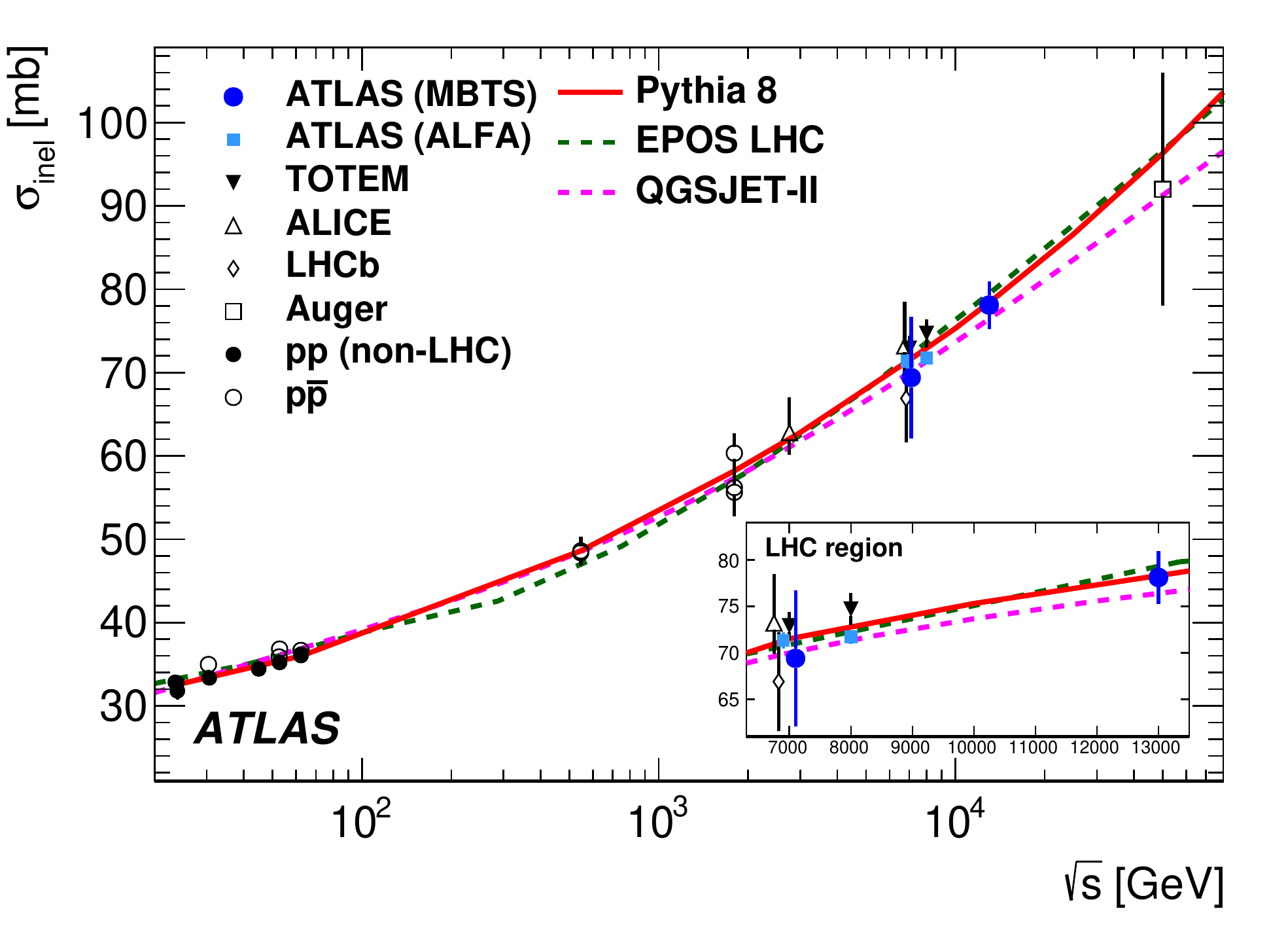}
\includegraphics[height=2in]{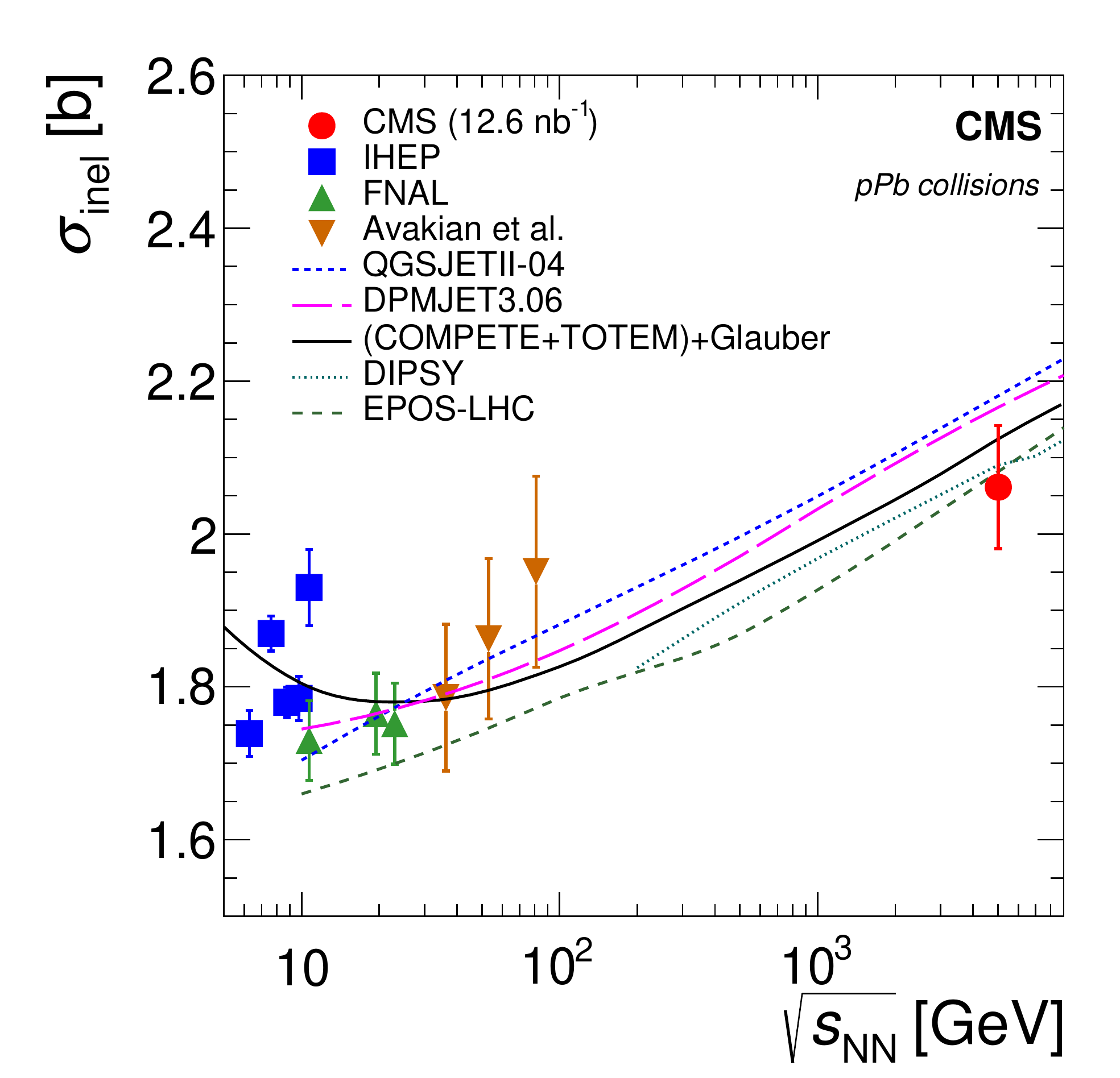}
\caption{ Left: The inelastic proton-proton cross section versus
  $\sqrt{s}$. The LHC data points are slightly shifted horizontally
  for better visibility~\cite{ATLAS-inel}. Right: The inelastic
  proton-Pb cross section versus $\sqrt{s}$~\cite{CMS-inel}. }
\label{fig:inel}
\end{figure}

   Figure~\ref{fig:inel} shows the compilations of the measurements of
   the inelastic proton-proton  and proton-Pb cross
    sections as a function of the center-of-mass energy, up to
    $\sqrt{s}=$13 TeV, compared to
    the predictions of several MC generators.  A good agreement is observed between
    measurements and the model expectations.

\begin{figure}[htb]
\centering
\includegraphics[height=2.1in]{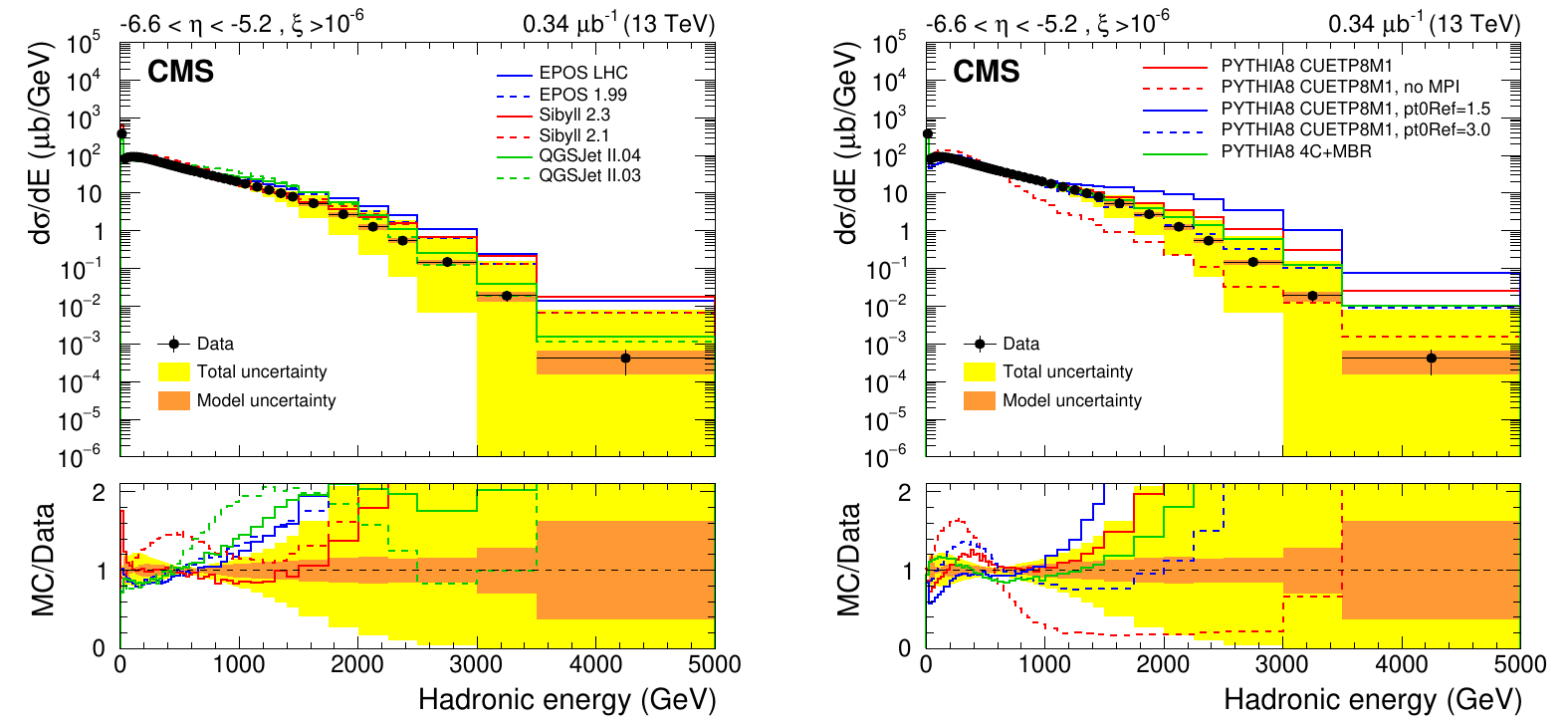}
\caption{ Differential cross section as a function of the hadronic
  energy in the region $-6.6<\eta<-5.2$ for events with
  $\xi>10^{-6}$. The left panel shows the data compared to MC event
  generators mostly developed for cosmic ray induced air showers,
  and the right panel to different PYTHIA 8 tunes~\cite{CMS-fwd-energy}. }
\label{fig:fwd_energy}
\end{figure}

The electromagnetic, hadronic, and total energy spectra of particles
produced at very forward pseudorapidities ($-6.6<\eta<-5.2$) have been
measured with the CASTOR calorimeter of the CMS experiment in
proton-proton collisions at a $\sqrt{s}=$13 TeV. The
experimental distributions, fully corrected for detector effects, are
compared to the predictions of various Monte Carlo event generators
commonly used in high energy cosmic ray physics (EPOS, QGSJetII, and
Sibyll), and those of different tunes of PYTHIA 8 (Figure~\ref{fig:fwd_energy}). None of the generators considered describe all features seen in the data. The present measurements are particularly sensitive to the modeling of multiparton interactions (MPI) that dominate particle production in the underlying event at forward rapidities in pp collisions. Event generators developed for modeling high energy cosmic ray air showers, tuned to LHC measurements at 0.9, 7, and 8 TeV, agree better with the present data than those tuned to Tevatron results alone. This is especially true for QGSJetII and Sibyll. However, all these models underestimate the muon production rate in extensive air showers because of their inaccurate description of the hadronic shower component~\cite{CMS-fwd-energy}. 

\section{Inclusive charged particle spectra in pp collisions }

\begin{figure}[htb]
\centering
\includegraphics[height=2.5in]{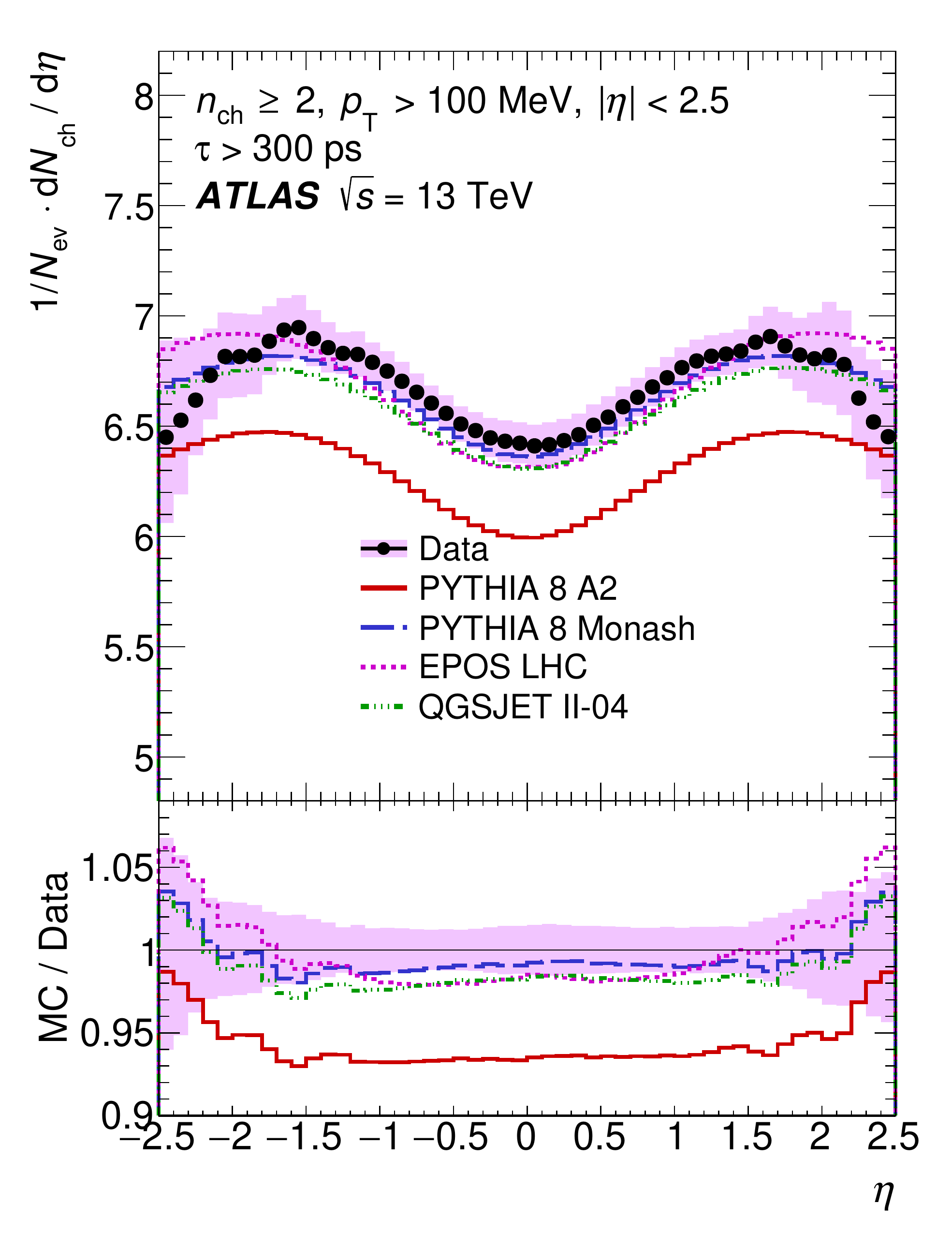}
\includegraphics[height=2.5in]{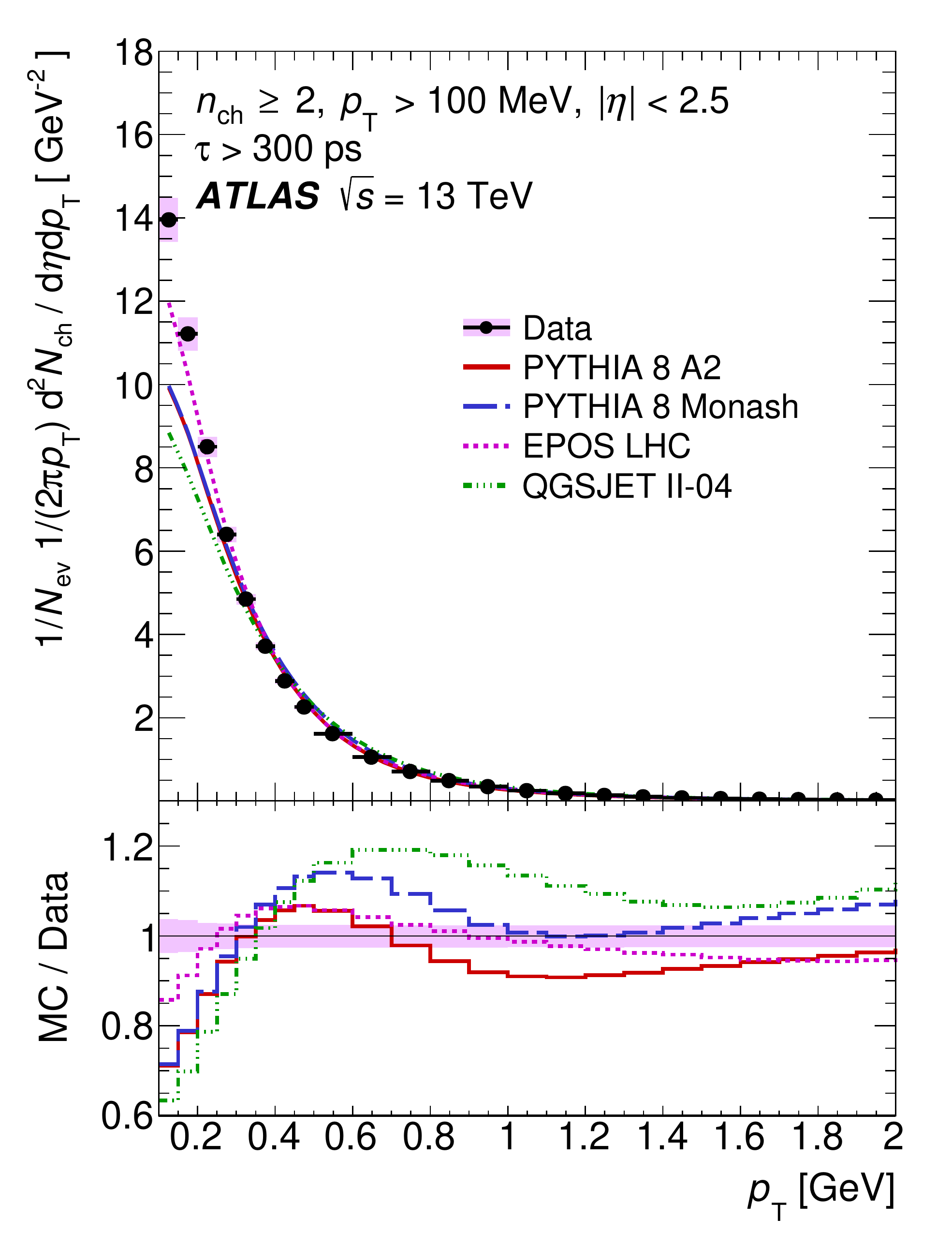}
\includegraphics[height=2.5in]{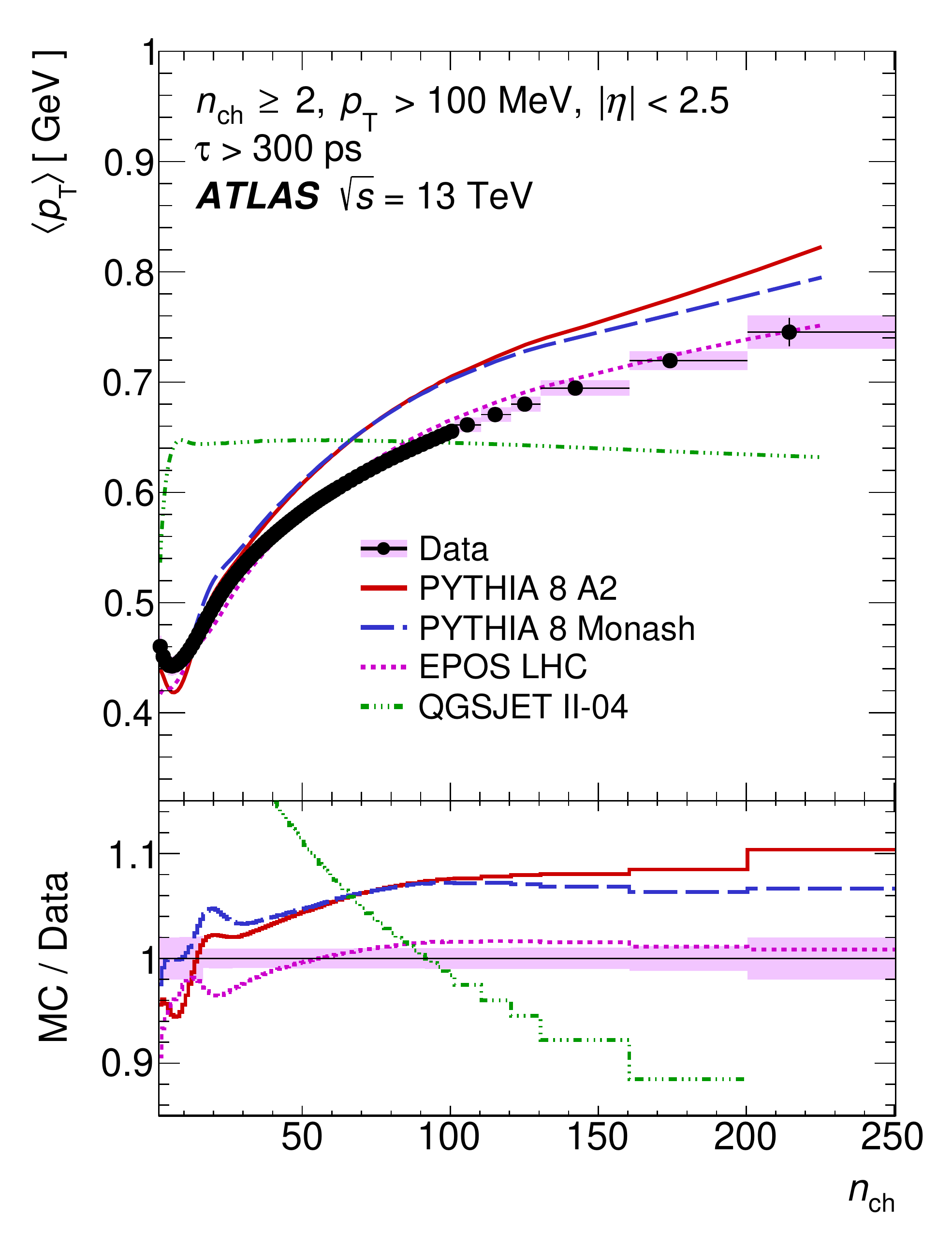}
\caption{ Primary charged-particle multiplicities as a function of (a)
  pseudorapidity and (b) transverse momentum, (c)  the mean transverse
  momentum ⟨ p$_{\rm T}$ ⟩ versus multiplicity for events with at
  least two primary charged particles with p$_{\rm T}>$100 MeV and
  $|\eta|<$2.5, each with a lifetime $\tau >$ 300 ps. The vertical bars represent the statistical uncertainties, while the shaded areas show statistical and systematic uncertainties added in quadrature~\cite{ATLAS-incl13}. }
\label{fig:inclusive_charged}
\end{figure}

  Figure~\ref{fig:inclusive_charged} shows the comparison of the
  inclusive charged-particle spectra measured by
  ATLAS~\cite{ATLAS-incl13}  at $\sqrt{s}=$ 13 TeV with the predictions of
  PYTHIA8 (A2 and Monash tune), EPOS LHC and QGSJET II-04 event
  generators. Overall, the data are best described by EPOS. The
  QSGJET-II generator, which has no model for colour coherence
  effects,  describes the $p_{\rm T}$ dependence poorly.

\subsection{Underlying event spectra in pp collisions }
\begin{figure}[htb]
\centering
\includegraphics[height=2.1in]{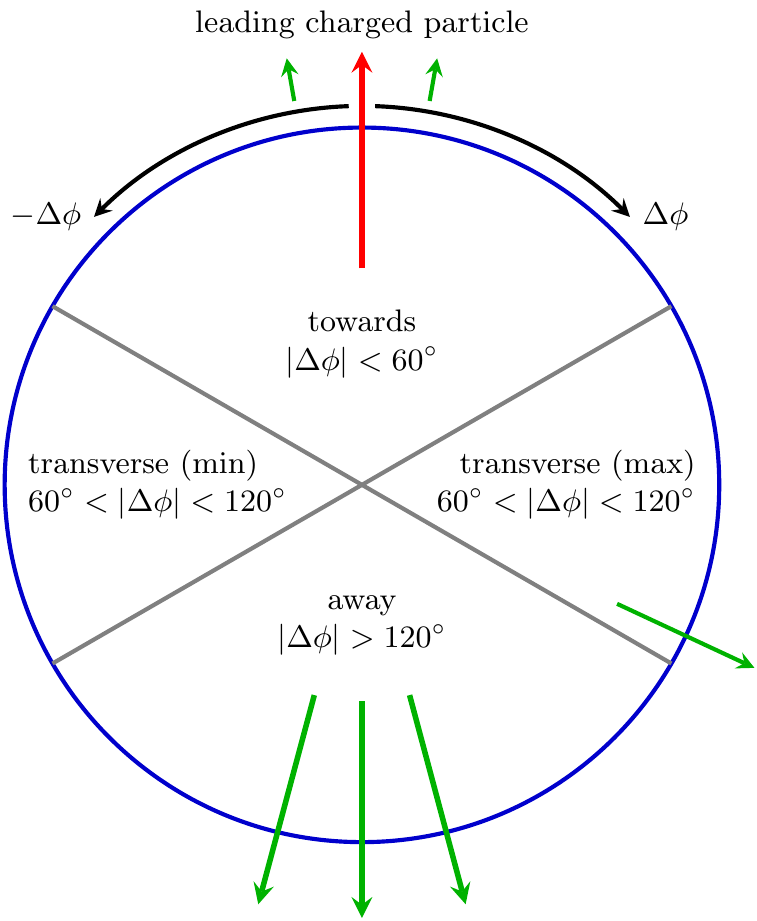}
\hspace{1in}\includegraphics[height=2.5in]{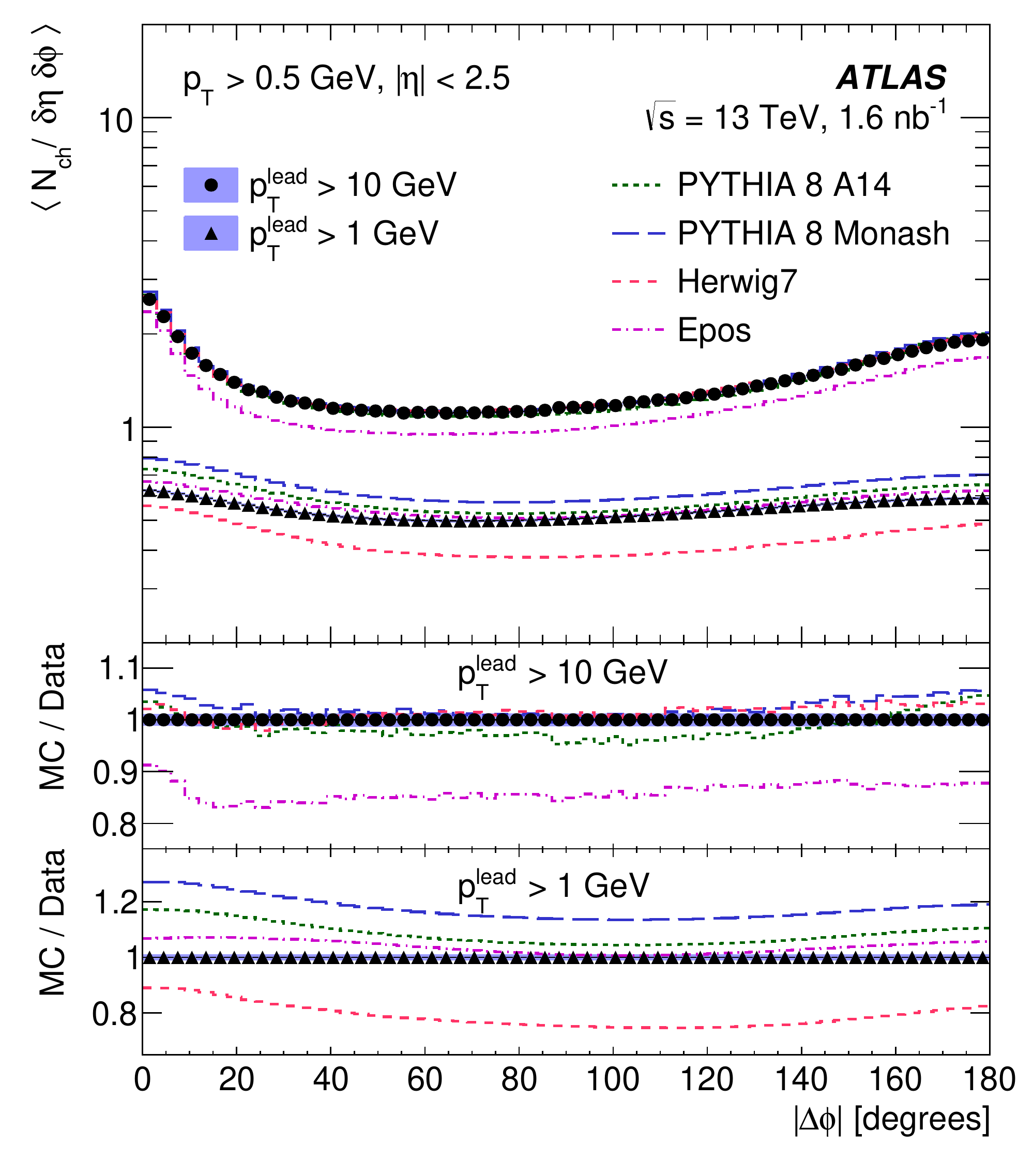}
\caption{Left: Definition of regions in the azimuthal angle with
  respect to the leading (highest-p$_{\rm T}$) charged particle, with
  arrows representing particles associated with the hard scattering
  process and the leading charged particle highlighted in red. Right:
  Distribution of mean charged-particle density as a function of 
 $|\Delta\phi|$ (with respect to the leading charged particle) for
 p$_{\rm T}^{\rm lead} >$ 1GeV and p$_{\rm T}^{\rm lead} >$ 10GeV separately, with comparisons to MC generator models~\cite{ATLAS-ue}. }
\label{fig:ue-yield}
\end{figure}

 Charged-particle distributions sensitive to the underlying event have
 been measured by the
ATLAS detector in low-luminosity proton-proton collisions at
$\sqrt{s}=$ 13 TeV~\cite{ATLAS-ue}. 
 The angular distribution of energy and particle flows with respect to the charged particle
with highest transverse momentum ( the leading particle) is defined as
shown in Figure~\ref{fig:ue-yield} (left). Figure~\ref{fig:ue-yield}
shows the measurement of the average charged-particle density for the
p$_{\rm T}^{\rm lead}>$1 GeV and p$_{\rm T}^{\rm lead}>$10 GeV. The results are compared
to the predictions of various Monte Carlo event generators, experimentally establishing the
level of underlying-event activity at LHC Run~2 energies and providing
inputs for the 
development of event generator modelling. The current models in use for UE modelling typically
describe this data to 5\% accuracy, compared with data uncertainties of less than 1\%.

The EPOS MC generator, specialised for simulation of inclusive soft QCD processes, displays particularly
discrepant features as the p$_{\rm T}^{\rm lead}$ scale increases, casting doubt on its suitability for modelling LHC multiple
pp interactions despite currently providing the best description of
minimum-bias data.  Further information about MC generator tunes can
be found in~\cite{CMS-ue}. 

\clearpage 
\section{Identified particle spectra}

\begin{figure}[htb]
\begin{minipage}{0.5\textwidth}
\centering
\includegraphics[height=3.5in]{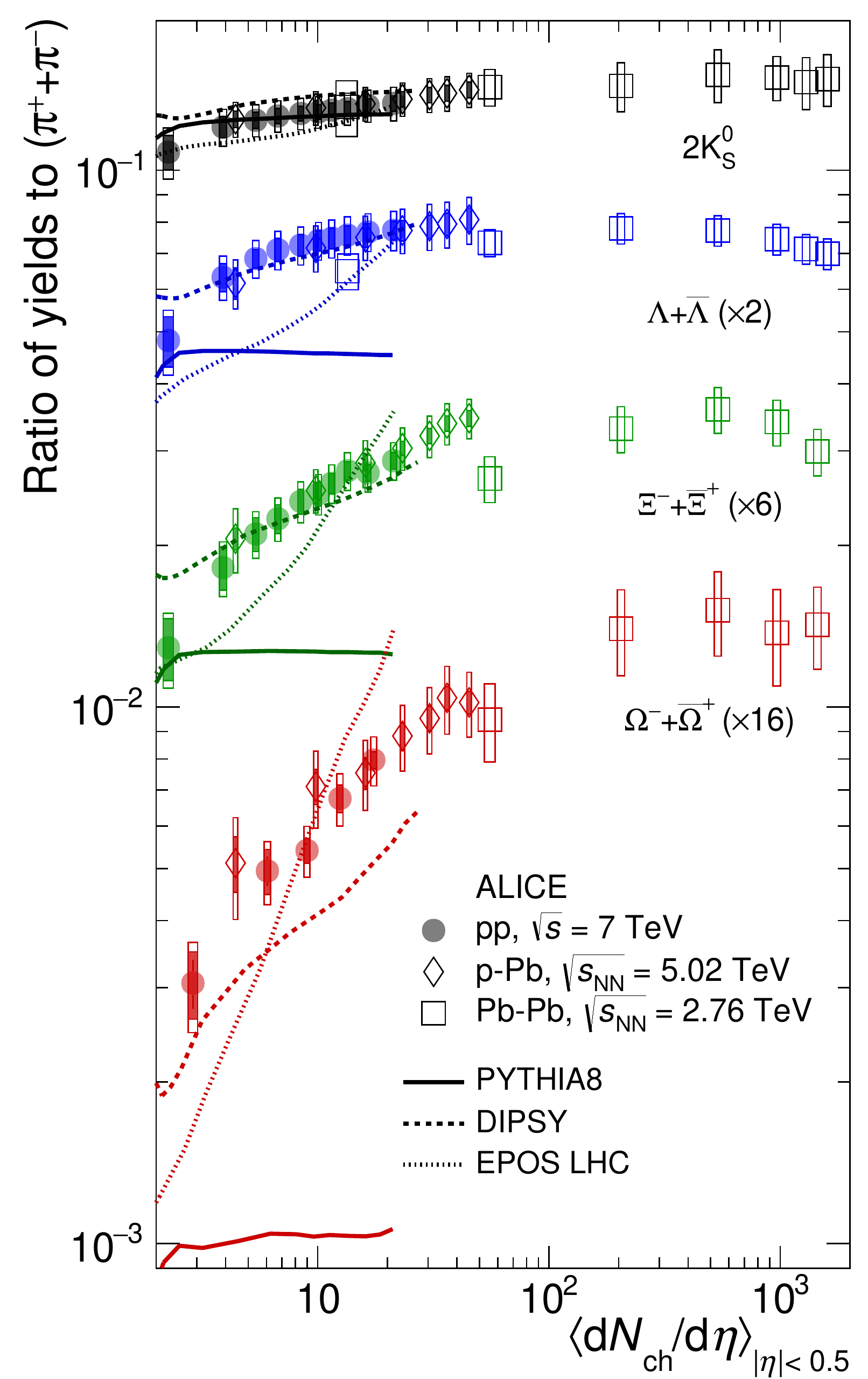}
\end{minipage}%
\begin{minipage}{0.5\textwidth}
\centering
\includegraphics[height=2.in]{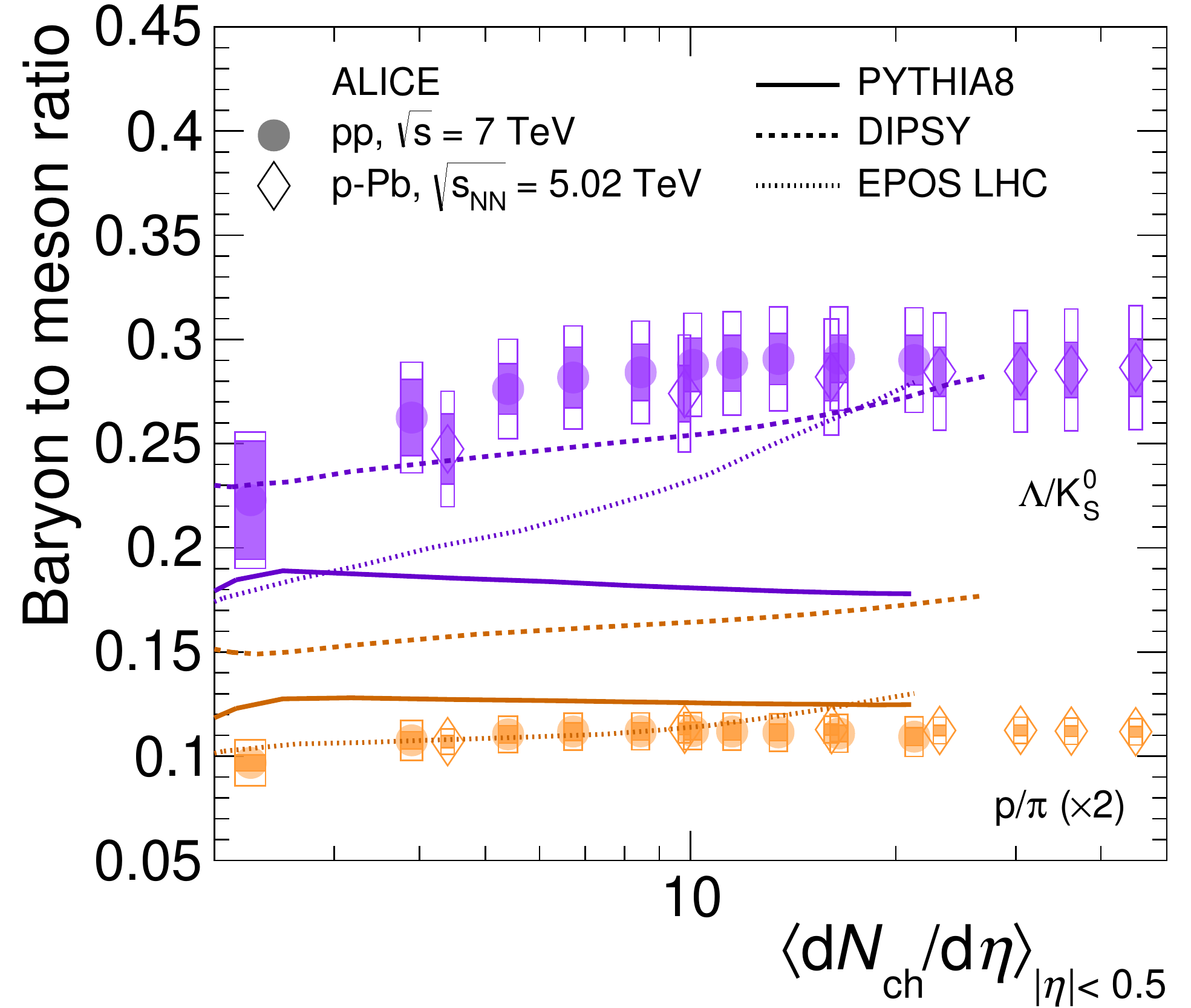}
\includegraphics[height=2.in]{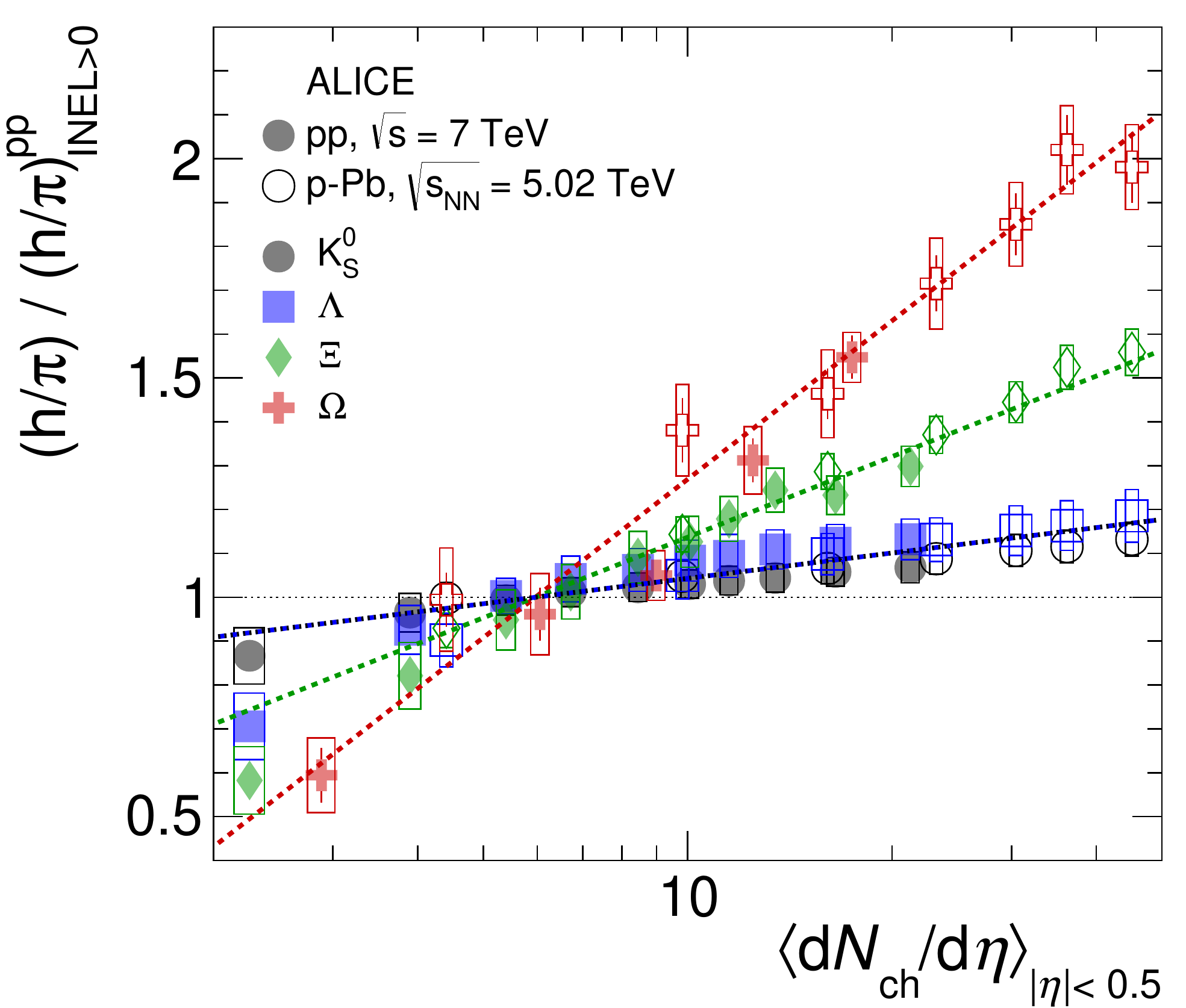}
\end{minipage}
\caption{Left: p$_{\rm T}$-integrated relative yields w.r.t. the pion 
  ($\pi^+$+$\pi^−$)  production as a function of charged-particle density measured
  in $|\eta|<$0.5. The error bars show the statistical uncertainty,
  whereas the empty and dark-shaded boxes show the total systematic
  uncertainty and the contribution uncorrelated across multiplicity
  bins, respectively. Top right: Ratio of baryon to meson yield
  compared with MC predictions. Bottom right: Phenomenological fit of
  the charge multiplicity dependence of the relative yield w.r.t. the
  pion production~\cite{ALICE-strange}. }
\label{fig:alice-strange}
\end{figure}
 ALICE Collaboration presented the first observation of strangeness
 enhancement in high-multiplicity pp collisions~\cite{ALICE-strange}.
 The measurement shows that the integrated yields of strange and
 multi-strange particles relative to pions increases significantly
 with the event charged-particle multiplicity
 (Figure~\ref{fig:alice-strange}, left), and that they are in remarkable
 agreement with p-Pb collision results  indicating that the phenomenon 
 is related to the final system created in the collision. In
 high-multiplicity events strangeness production reaches values
 similar to those observed in Pb-Pb collisions. The multiplicity
 dependence is best described by the DIPSY model which however
 overestimates the yield of protons w.r.t. the pion production
 (Figure~\ref{fig:alice-strange}, top right).

\section{Particle correlations}

\begin{figure}[htb]
\centering
\includegraphics[height=2.2in]{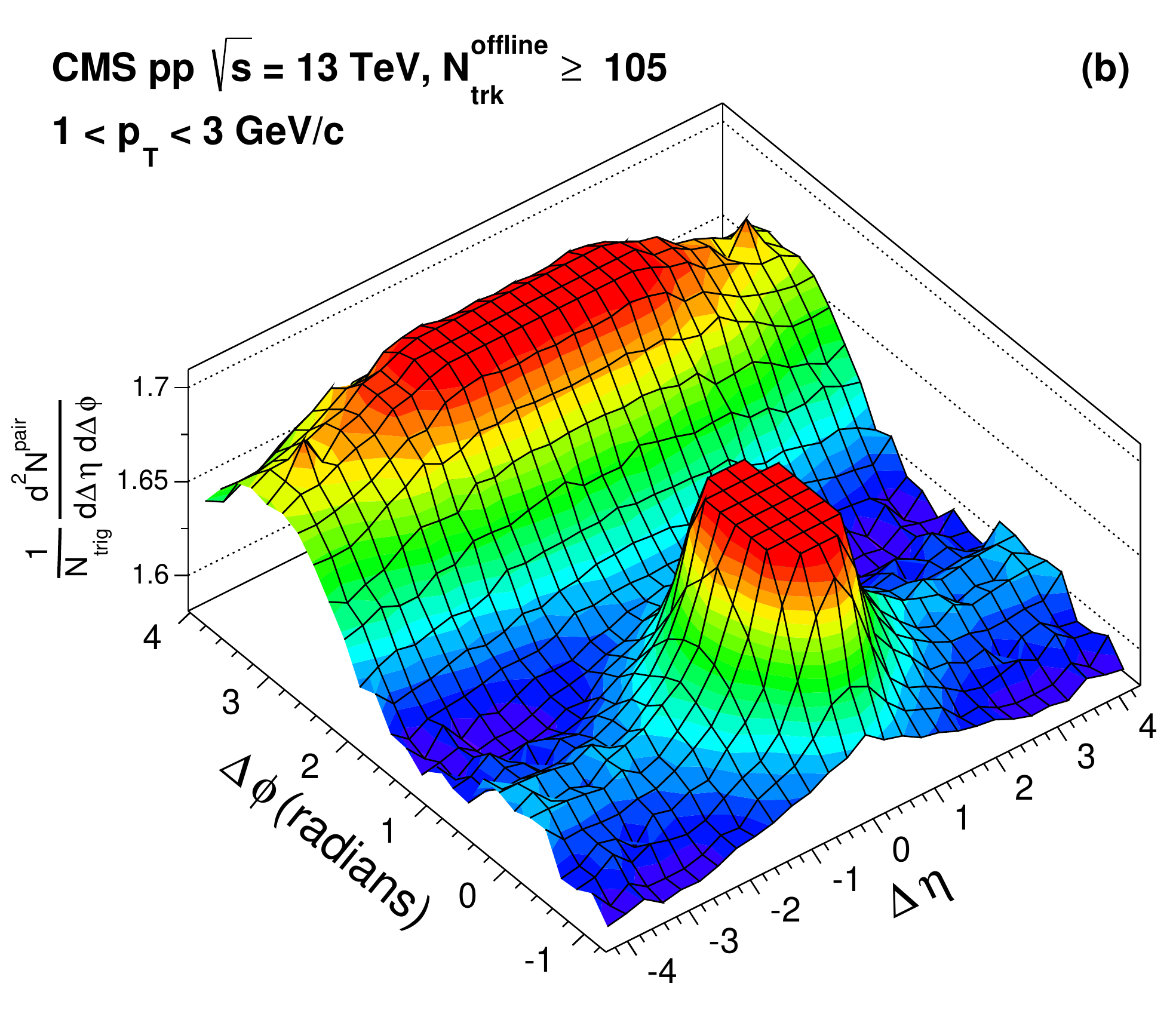}
\hspace{0.5in}\includegraphics[height=2in]{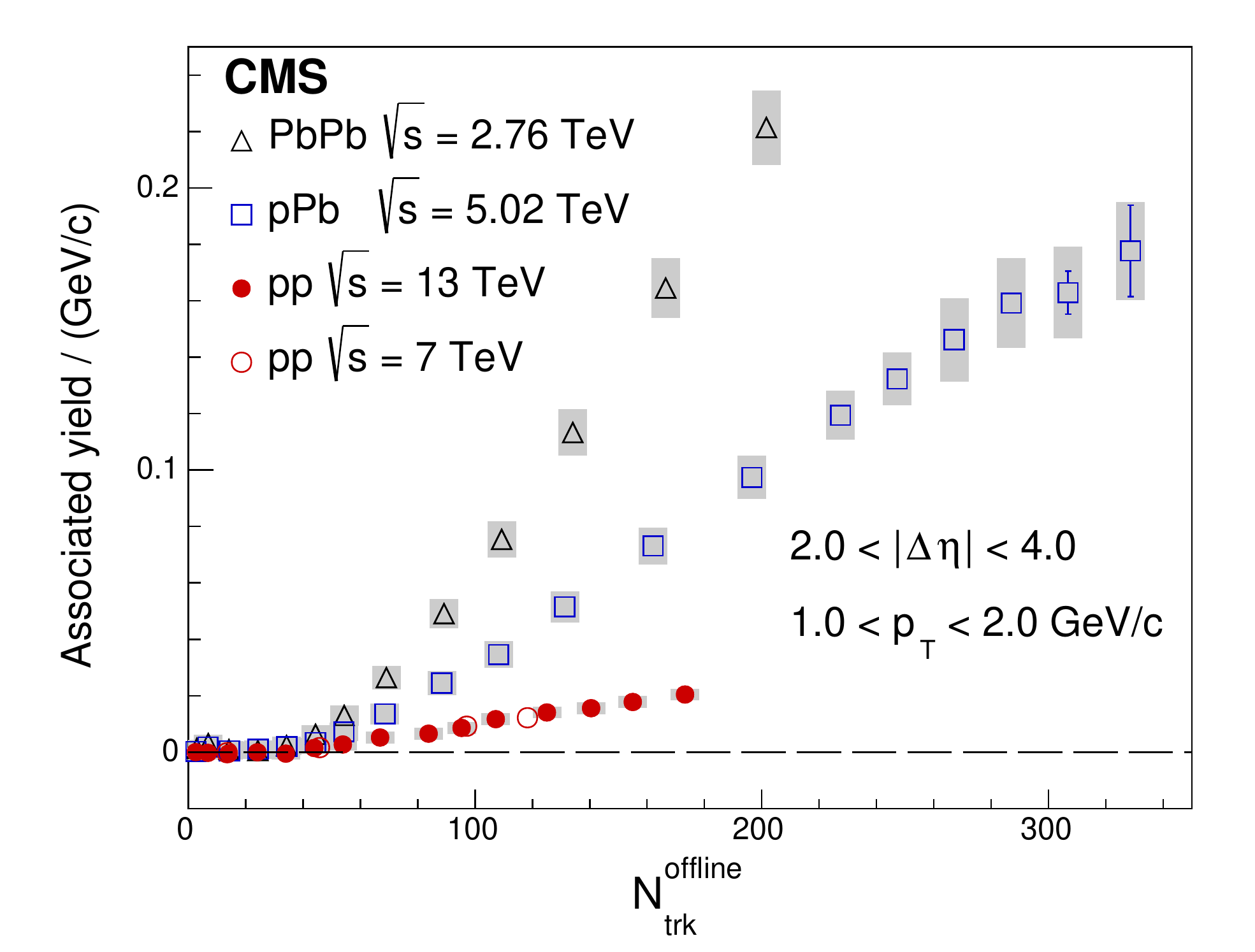}
\caption{ Left: The 2-dim ($\Delta\eta,\Delta\Phi$) two-particle
  correlation functions for pairs of charged particles both in the
  range 1 $<p_{\rm T}<$ 3 GeV/c, for charged multiplicity above
  105. The sharp peaks from jet correlations around (Δη,Δϕ)=(0,0) are
  truncated to better illustrate the long-range correlations. Right:
  Associated yields of long-range near-side two-particle correlations,
  averaged over 2$ <|\Delta\eta|<$ 4. The error bars correspond to the statistical uncertainties, while the shaded areas denote the systematic uncertainties. Note that there are PbPb points above the upper vertical scale, which are not shown for clarity~\cite{CMS-ridge}.}
\label{fig:cms-ridge}
\end{figure}

   The particle yields corresponding to the near-side long-range
   correlations are measured by the CMS Collaboration~\cite{CMS-ridge} in
 pp collisions at $\sqrt{s}=$ 13 TeV and compared with
 previous measurement at $\sqrt{s}=$7 TeV and with the pPb and PbPb
 data (Figure~\ref{fig:cms-ridge}).  The observed yields have
 approximately linear dependence on the event charge multiplicity for
 multiplicity above $\sim$40.
 Further measurements of the
 near-side correlations can be found in~\cite{ALICE-ridge},
 \cite{ATLAS-ridge}, \cite{LHCb-ridge}. A study of long-range
 correlations with the subevent cumulant method finds a close
 similarity between the multiplicity dependence of  the number of
 sources in pp and pPb collisions \cite{ATLAS-ridge-subevent}.

\begin{figure}[htb]
\centering
\includegraphics[height=2.1in]{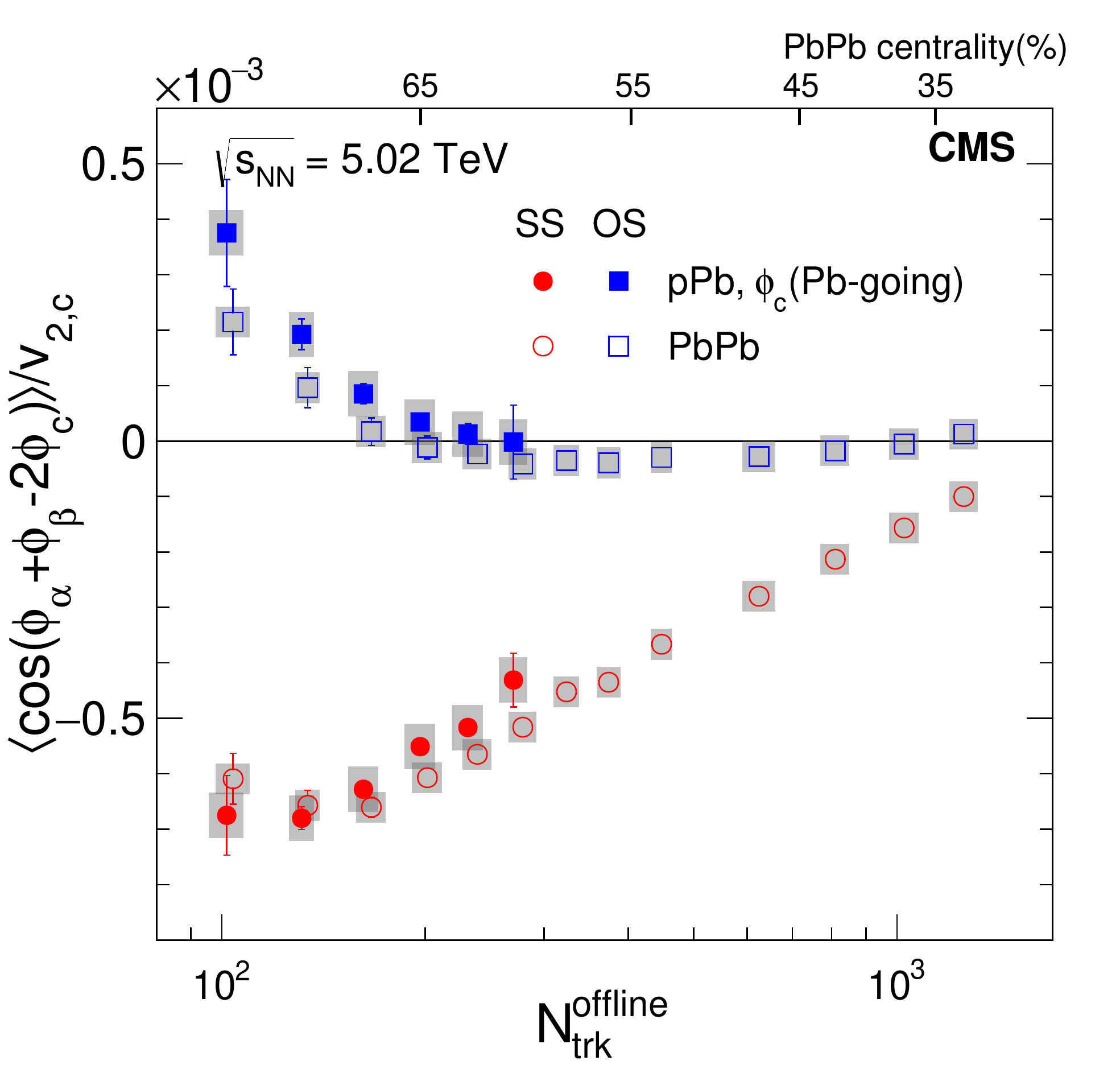}
\includegraphics[height=2.1in]{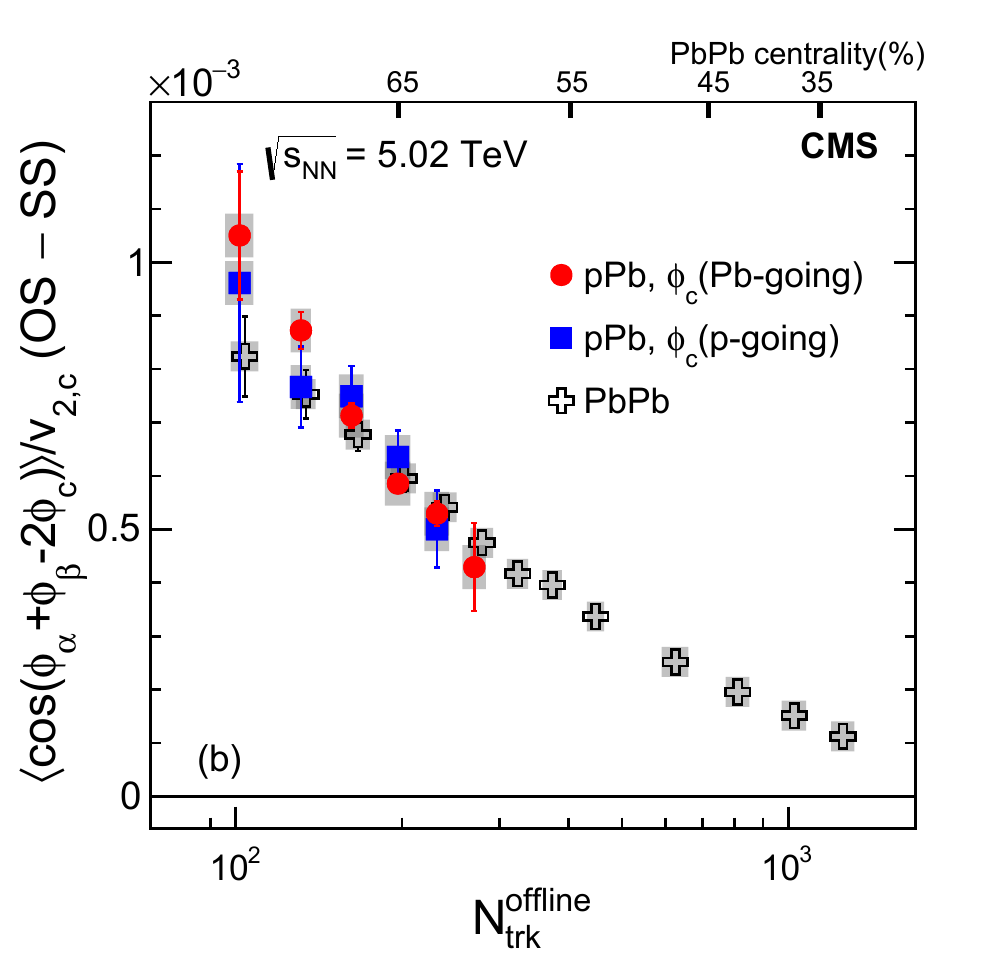}
\caption{ Left: the same-sign (SS) and opposite-sign (OS) three-particle
  correlators averaged over $|\eta_{\alpha}−\eta_{\beta}|<$ 1.6. 
  Right:  their difference  as a function of charged multiplicity
  in pPb and PbPb collisions~\cite{CMS-chaz}.}
\label{fig:corr_chaz}
\end{figure}

Charge-dependent azimuthal correlations of the same- and opposite-sign
particle pairs with respect to the second-order event plane have been
measured in pPb and PbPb collisions at $\sqrt{s_{NN}}=$ 5.02 TeV by
the CMS experiment at the LHC~\cite{CMS-chaz}. Correlations are extracted via a
three particle correlator as functions of the $\Delta\eta$ and the
charged-particle multiplicity of the event. The differences between
the same- and the opposite-sign particle pairs agree for pPb and PbPb collisions
(Figure~\ref{fig:corr_chaz}), possibly indicating a common underlying 
mechanism that generates the observed correlation. These results
challenge the CME interpretation
for the observed charge-dependent azimuthal correlations in nucleus-nucleus collisions at RHIC and the LHC. 

\begin{figure}[htb]
\centering
\includegraphics[height=2in]{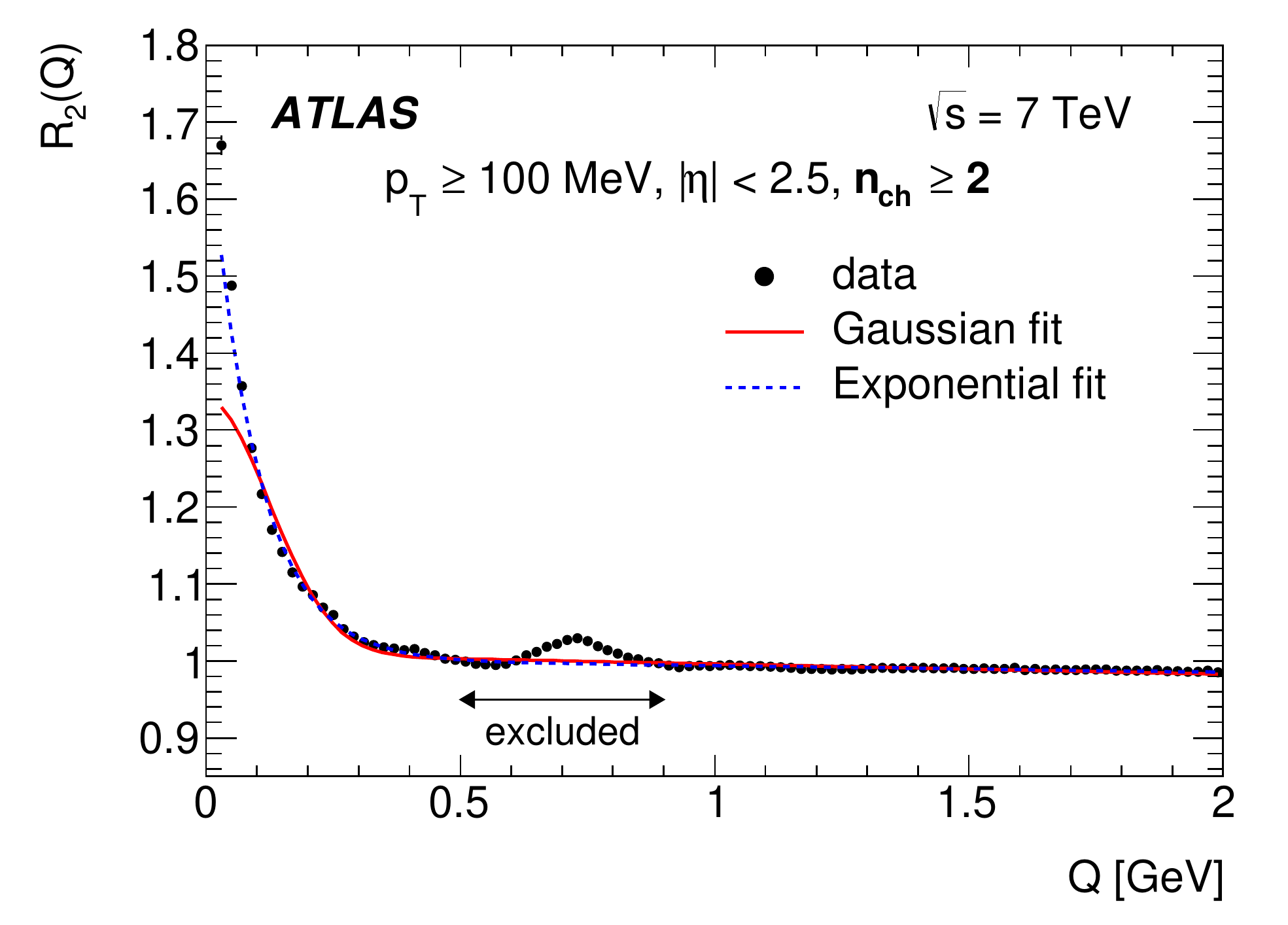}
\includegraphics[height=2in]{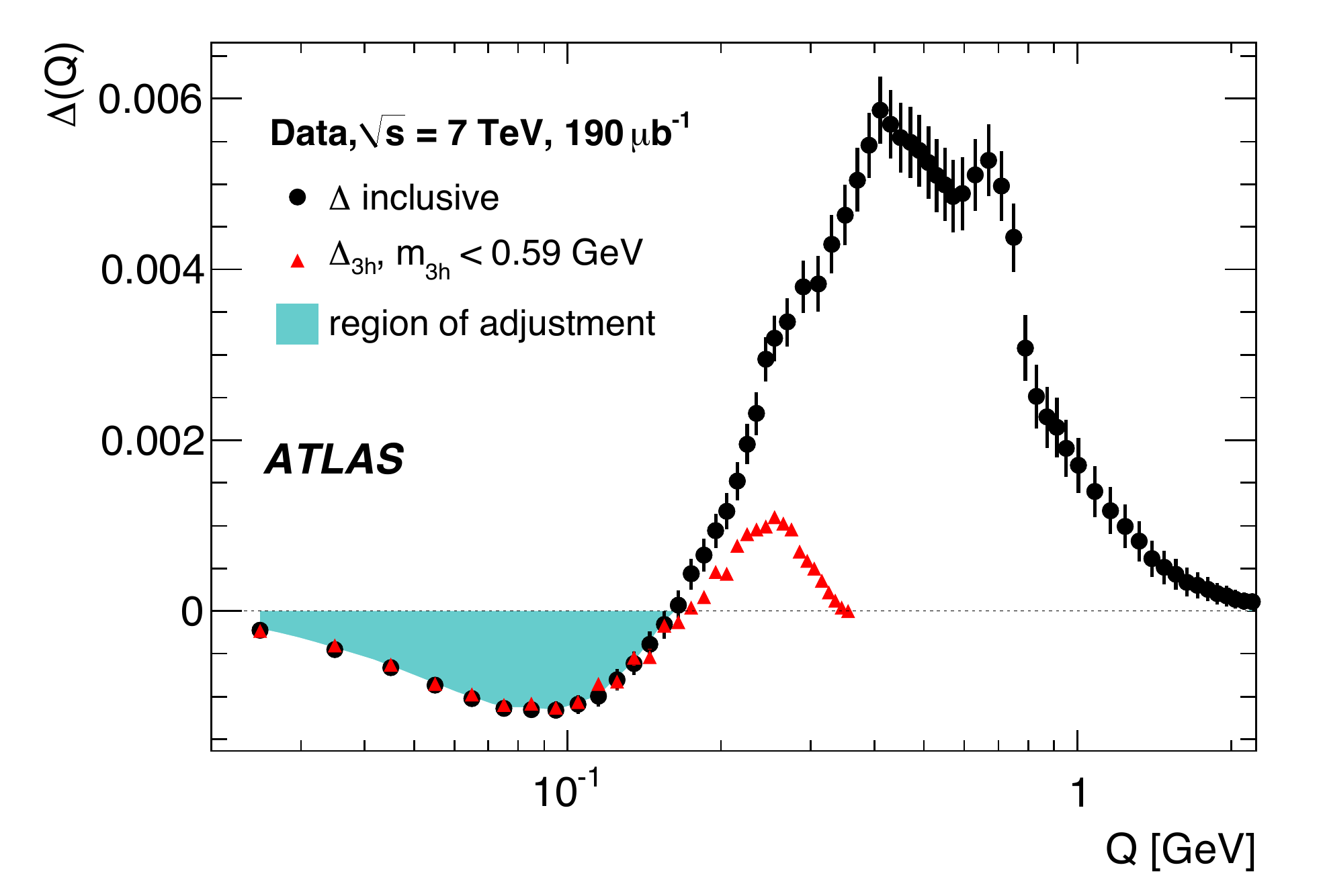}
\caption{  The enhanced production of same-sign particle pairs
  analysed (left) as a signature of the Bose-Einstein
  interference~\cite{ATLAS-BE}, via the ratio of same-sign and
  opposite-sign pair distributions,  and (right) as
  a secondary effect of correlations between adjacent hadrons in the
  study of ordered hadron chains~\cite{ATLAS-chains}, via the
  subtraction of same-sign pair distribution from the opposite-sign pair distribution. }
\label{fig:chains}
\end{figure}

  Another example of a measurement challenging the conventional interpretation is the study of
  ordered hadron chains~\cite{ATLAS-chains}. The measurement isolates the source of
  two-particle correlations at low four-momentum difference with help
  of charge-ordered hadron triplets with minimized mass (Figure~\ref{fig:chains}). The
  correlations are shown to be compatible with the signature of
  helical QCD string fragmenting into a set of ground-state pions. The measurement offers an alternative
  explanation of the enhanced production of close like-sign pairs,
  traditionally attributed to the Bose-Einstein
  effect~\cite{ATLAS-BE}.  
 
\begin{figure}[htb]
\centering
\includegraphics[height=2.2in]{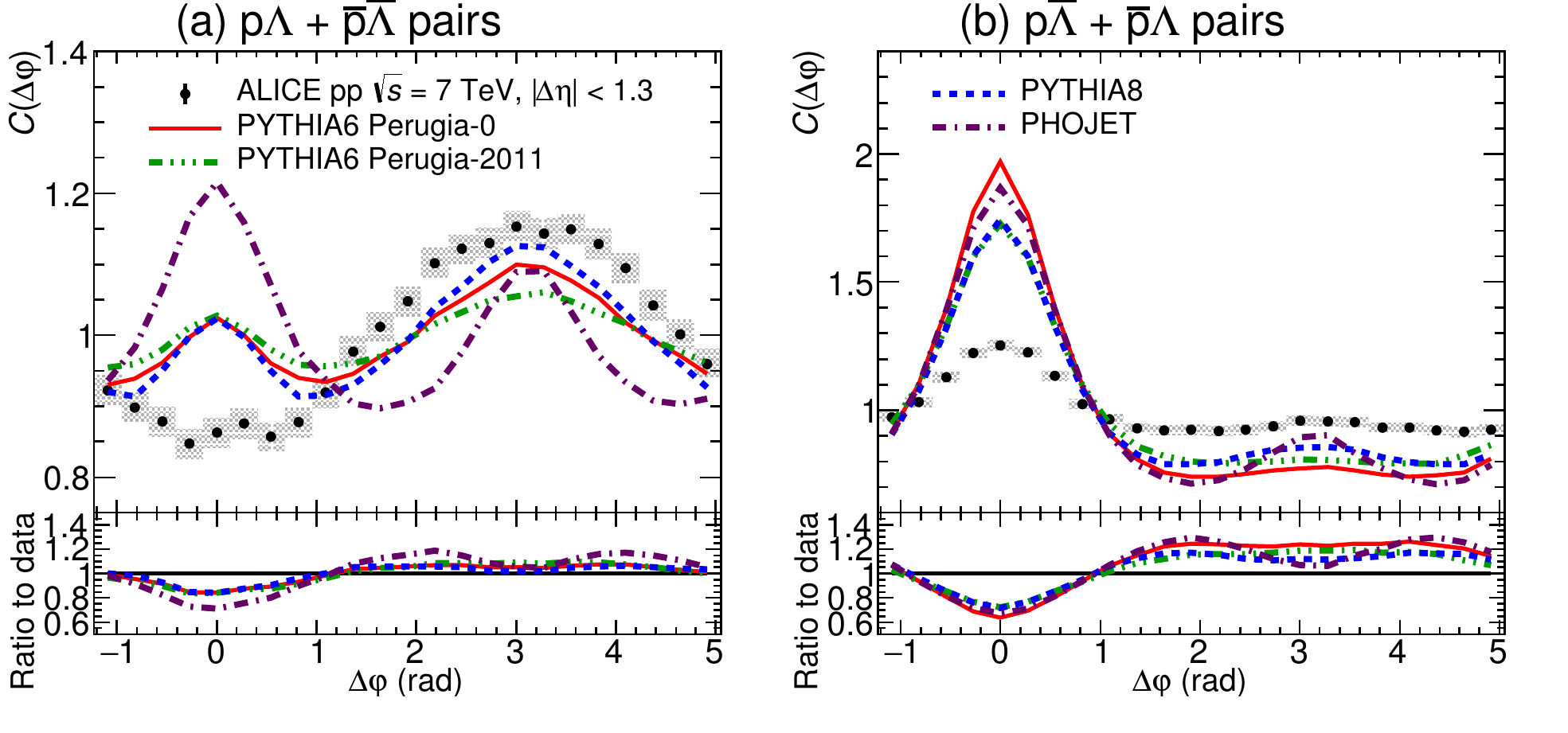}
\caption{ Example of anti-correlations observed for baryon pairs. $\Delta\eta$
integrated projection of correlation functions for combined pairs of (left) $p\Lambda$+$\bar{p}\bar{\Lambda}$
and (right) $p\bar{\Lambda}$+$\bar{p}\Lambda$ 
obtained from ALICE pp collision data and four Monte Carlo models (PYTHIA6 Perugia-0, PYTHIA6 Perugia-
2011, PYTHIA8 Monash, PHOJET) at $\sqrt{s}=$7 TeV~\cite{ALICE-id-corr}.}
\label{fig:baryon_corr}
\end{figure}

Two-particle angular correlations of identified particles were
measured in proton-proton collisions at $\sqrt{s}$=7 TeV by ALICE Collaboration~\cite{ALICE-id-corr}. 
The analysis was carried out for pions, kaons, protons, and lambdas,
for all particle/anti-particle combinations in the pair. Data for
mesons exhibit an expected peak dominated by effects associated with
mini-jets and are well reproduced by general purpose Monte Carlo
generators. However, for baryon--baryon and anti-baryon--anti-baryon
pairs, where both particles have the same baryon number, a near-side
anti-correlation structure is observed instead of a peak (Figure ~\ref{fig:baryon_corr}). This effect is interpreted in the context of baryon production mechanisms in the fragmentation process. It currently presents a challenge to Monte Carlo models and its origin remains an open question.

%


\section{Conclusions}
      
    A large variety of ongoing measurements bring new insights and
  constraints to our understanding of the hadron production.   There
  are still quite a few unresolved problems, but we possess a wealth
  of data which will gradually help us to shed light on rules
  governing the hadronization, which may explain many similarities
  observed between pp, pPb and PbPb collisions, and the soft
  interactions in general.
 
  Many thanks to all contributing authors !


\end{document}